\documentclass[%
 reprint,
superscriptaddress,
showkeys,
 amsmath,amssymb,
 aps,
pre, 
]{revtex4-2}

\usepackage{graphicx}% Include figure files
\usepackage[caption=false]{subfig} %?

\usepackage{dcolumn}% Align table columns on decimal point
\usepackage{bm}% bold math
\newcommand{\ee}{\end{equation}} 
\newcommand{\be}{\begin{equation}}

\usepackage[mathscr]{euscript}  
\usepackage{xcolor}  

\usepackage{tcolorbox}
\usepackage{comment}
\usepackage{float}
\usepackage{soul}

\makeatletter
\newsavebox{\@brx}
\newcommand{\llangle}[1][]{\savebox{\@brx}{\(\m@th{#1\langle}\)}%
  \mathopen{\copy\@brx\kern-0.5\wd\@brx\usebox{\@brx}}}
\newcommand{\rrangle}[1][]{\savebox{\@brx}{\(\m@th{#1\rangle}\)}%
  \mathclose{\copy\@brx\kern-0.5\wd\@brx\usebox{\@brx}}}
\makeatother

\begin{document}

\preprint{APS/123-QED}
%\title{ Directional switching of two-dimensional active swarms: the role of long-range interactions and delay time}
%\title{Collective directional switching in two-dimensional active matter: a \\ competition of time scales and the role of confinement}
%\title{Novel collective states of active matter with directional switching: the role of interaction relaxation times and confinement}
 
\title{Collective states of active matter with stochastic reversals: emergent\\ chiral states and spontaneous current switching}

\author{Kristian St\o{}levik Olsen}
\affiliation{Nordita, Royal Institute of Technology and Stockholm University,
Hannes Alfvéns väg 12, 23, SE-106 91 Stockholm, Sweden\\}

\author{Luiza Angheluta}
\affiliation{The Njord Center, Department of Physics, University of Oslo, Blindern, 0316 Oslo, Norway\\}

\author{Eirik Grude Flekk\o{}y}
\affiliation{PoreLab, the Njord Center, Department of Physics, University of Oslo, Blindern, 0316 Oslo, Norway\\}

\date{\today}

\begin{abstract}
We study collective dynamical behavior of active particles with topological interactions and directional reversals. Surprising phenomena are shown to emerge as the interaction relaxation time is varied relative to the reversal rate, such as spontaneous formation of collective chiral states due to phase synchronization, and collective directional reversals in the presence of confinement. The results have a direct relevance to  modelling and understanding collective reversals and synchronization phenomena in active matter.  
\end{abstract}

\pacs{Valid PACS appear here} % PACS, the Physics and Astronomy  Classification Scheme.
\keywords{Active matter; Collective phenomena; Kuramoto model;}
\maketitle
%%%%%%%%%%%%%%%%%%%%%%%%%%

Collective behavior is ubiquitous in biological systems, ranging from collective migration of cells and bacteria to flocks of birds \cite{marchetti2013hydrodynamics}. Multistability is often present in realistic swarming behavior, where the system undergoes a series of transitions between different collective states, triggered by internal or external perturbations. Examples include the co-existence of translational flocking and milling states observed for example in fish \cite{tunstrom2013collective} or bi-stable clockwise and anti-clockwise states of swarming of locust in circular confinements or annuli \cite{yates2009inherent, buhl2006disorder}.

Understanding the role disorder plays in collective phenomena has become a central question in modern active matter research. The case of external quenched disorder has been studied rather thoroughly both for interacting and non-interacting active particles \cite{martinez2021active, martinez2018collective, duan2021breakdown, chardac2021emergence, reichhardt2021active, reichhardt2021clogging, martinez2020trapping, morin2017distortion, olsen2021active, alonso2019transport,  peruani2018cold,sandor2017dynamic,borba2020controlling, adhikary2021effect,chepizhko2015active, chepizhko2013optimal, pattanayak2019enhanced,rahmani2021topological}. More interesting perhaps is the role played by internal types of disorder and complexity, originating not from external factors but rather from the dynamics of the particles themselves. In the case of mobility of animals or microorganisms such intrinsic disorder corresponds for example to behavioral variability in a population. Recent studies include inhomogeneous chirality disorder throughout a population \cite{ventejou2021susceptibility}, speed inhomogeneity \cite{pattanayak2020speed} or the inclusion of dissenters \cite{yllanes2017many} or particles with contrarian tendencies \cite{bonilla2019contrarian}.  

Active matter in confinement has also gained more interest because it is representative in biological systems and has potential microfluidic applications. Surprising behaviors emergence under confinement, including accumulation phenomena near boundaries \cite{fily2014dynamics, fazli2021active, leoni2020surfing,caprini2018active}, non-trivial escape dynamics \cite{PhysRevResearch.2.043314, PhysRevE.102.042617} and oscillating collective motion  \cite{zhang2020oscillatory, liu2020oscillating}. In such bounded spaces, the collective states typically conforms to the symmetry of the confinement. 

In this paper, we report on how topological alignment interactions and individual reversals in systems of non-chiral active particles produce striking collective behaviors, such as global chiral states in open spaces and collective directional reversals under confinement. For this, we propose a minimal model with two dimensionless parameters and study its mean-field limit to predict the transition point from disordered to partially-ordered states. We use extensive numerical simulations to explore the bistable dynamics of partially-ordered states in both open space and under a confinement that breaks rotational symmetry.

\begin{figure}
    \centering
    \includegraphics[width = 8.3cm]{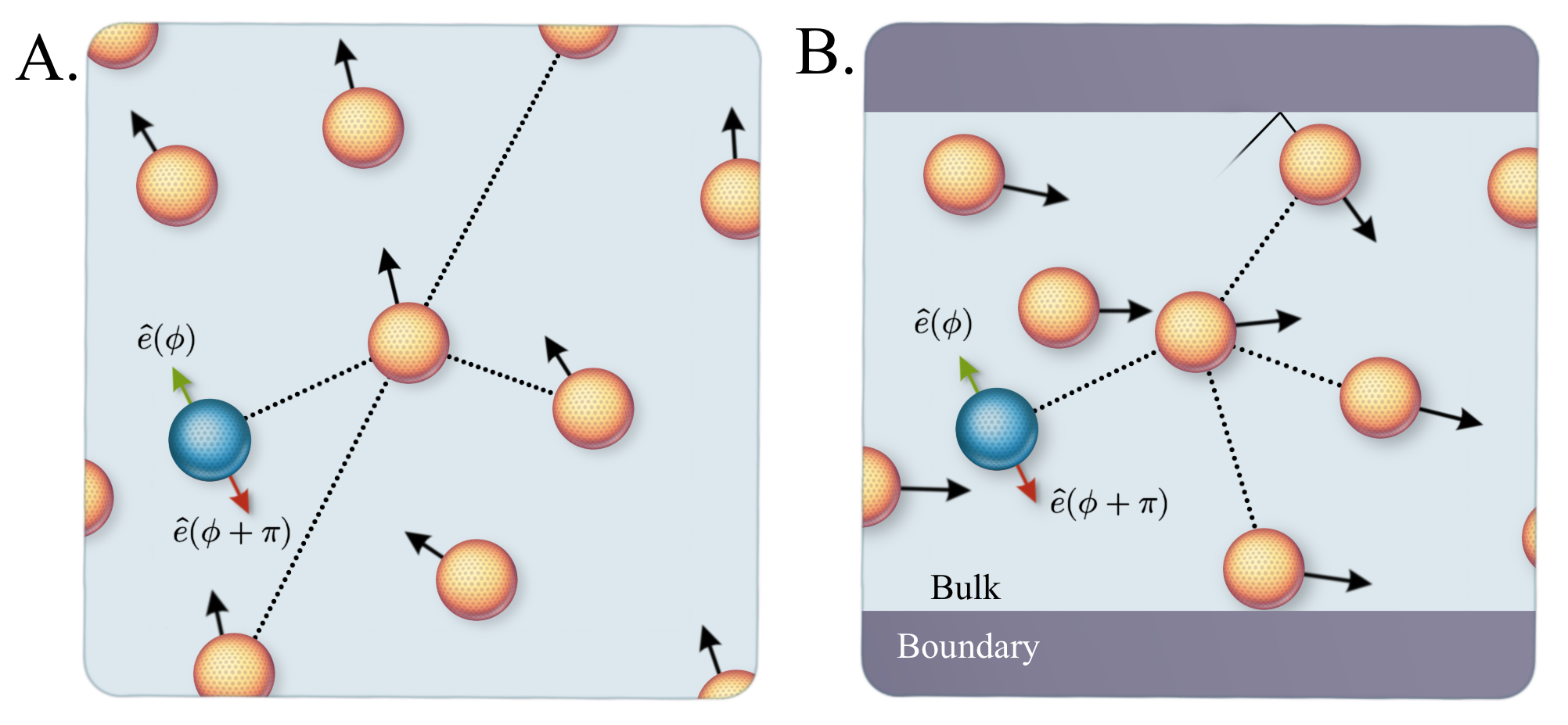}
    \caption{Sketch of the system under consideration. Self-propelled particles align through a Kuramoto interaction and undergo Poissonian stochastic reversals of their direction of motion (eg. blue particle). Interactions are topological, and each particle interacts with a fixed number $N_\text{int}$ of randomly chosen particles ($N_\text{int}=4$ in sketch). The case of open unbounded space (A) and confinement in the form of a channel (B)  is considered. }
    \label{fig:sketch}
\end{figure}

Crucial to the collective reversal phenomena observed in our model is the competition between the time-scale associated with single-particle reversals (SPR) and a relaxation time associated with the alignment interactions. Each active particle has an orientation $\varphi_i$ which evolves according to a Kuromoto equation, i.e. 
$ \dot \varphi_i = -\frac{1}{\tau_a} \left\langle \sin\left(\varphi_i-\varphi_j\right)\right \rangle_{\mathcal{I}_i}$ where $i = 1, 2, ..., N$ is the particle index. This is similar to the phase evolution of coupled oscillators, e.g.~\cite{strogatz2000kuramoto}, with a homogeneous coupling strength given here by the inverse of the relaxation time $\tau_a$. However, there is fundamental difference here in that our "oscillators" are moving in space in the direction of their phase and with a constant speed. Alignment is modelled using a simple topological interaction rule whereby particle $i$ is given a random set $\mathcal{I}_i$ of $N_\textbf{int}<N$ interaction partners. Such interactions correspond, for example, to non-local interactions mediated by sensory input in animal swarms~\cite{ballerini2008interaction, zumaya2018delay}. In addition, we induce single-particle reversals (SPR) as a stochastic switching modelled by a Poisson process, where the probability of reversal in a time window $\Delta t$ is $p_\text{rev} = e^{-\gamma \Delta t} \gamma \Delta t$, with rate parameter $\gamma$. Active matter systems with instantaneous Poissonian reversals have gained more attention in recent times both for interacting and non-interacting cases \cite{abbaspour2021effects, mahault2018self,santra2021active, PhysRevE.103.052608}. Such abrupt reversals are also observed in Nature, for example in some species of bacteria, including P. Putida and M. Xanthus \cite{pohl2017inferring, PhysRevE.104.L012601}. The reversal rate $\gamma$ sets the other timescale competing with alignment timescale. The ratio between them provides us with a dimensionless parameter $\mathcal{D} = (\gamma \tau_a)^{-1}$, which we will refer to as the delay number. Thus, in its dimensionless form, the evolution of the orientation angle $\varphi_i(t)$ reads as 
\begin{equation}\label{eq:stochastic}
    \dot \varphi_i = -\mathcal{D} \left\langle \sin\left(\varphi_i-\varphi_j\right)\right \rangle_{\mathcal{I}_i} + \pi \sum_\alpha \delta(t-t_\alpha^{(i)}).
\end{equation}
For simplicity, we neglect the rotational diffusion term in the limit where the timescale associated with the rotational meanderings is much larger than the other relevant timescales. 

\begin{figure}
    \centering
    \includegraphics[width = 8.7cm]{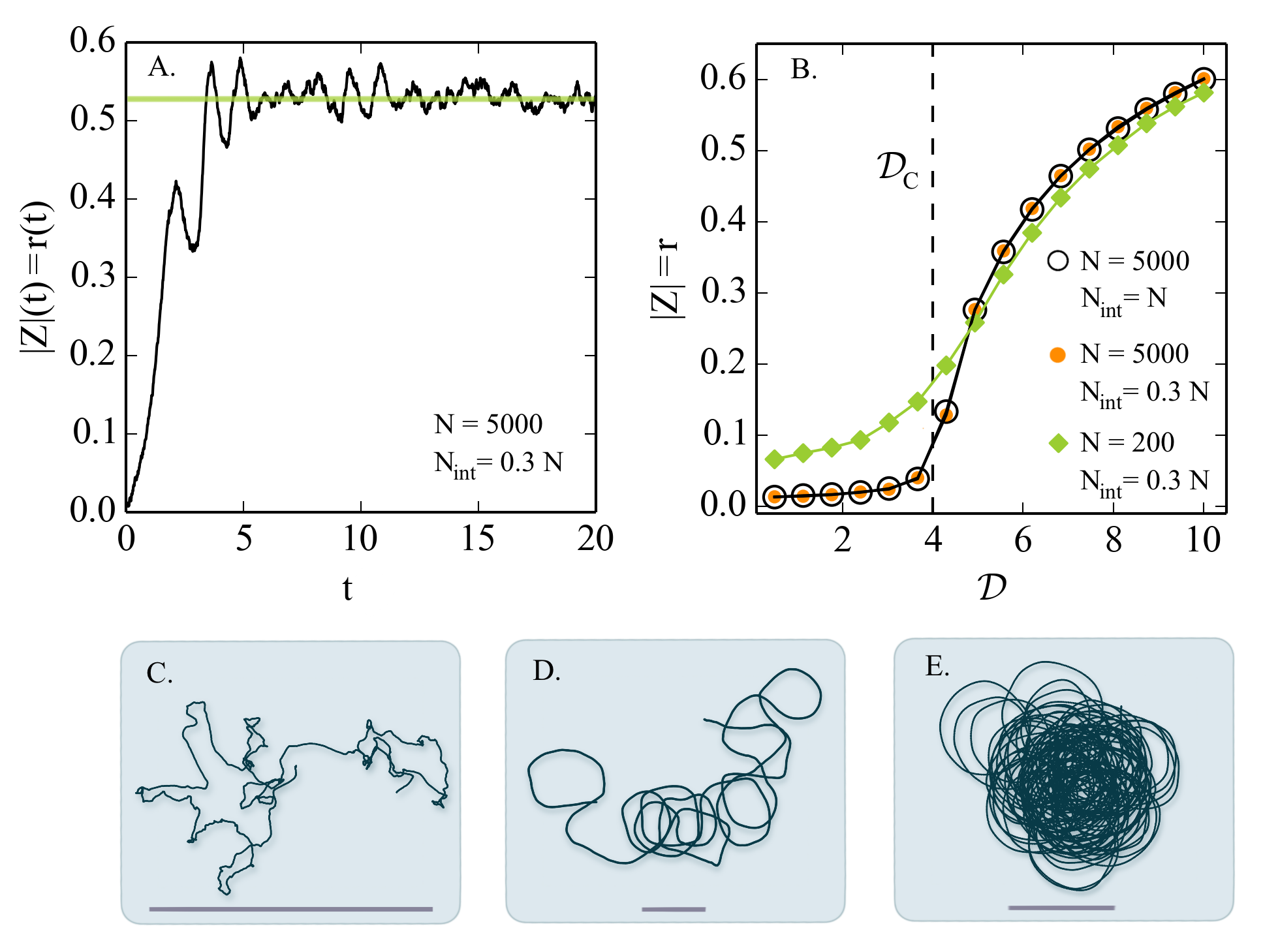}
    \caption{A. Time-series of the order parameter $|Z|(t)$ plateaus with small fluctuations around an asymptotic constant value $|Z_0|$. B) Asymptotic values of $|Z|$ as a function of the delay number $\mathcal{D}$, shows both for the mean-field limit $N_\text{int} = N-1$ and away from the mean-field limit. Dashed line indicates the predicted (mean-field) critical value of the delay number where order starts to form. Figures C-E shows center-of-mass trajectories for $N = 200$ , $N_\text{int} = 0.3 N$ (corresponding to the green curve in B.), with delay numbers $\mathcal{D} = 1,6,15$ respectively. Relative scale is shown by the bar in the bottom of figures C-E). $\text{Pe}_s =1$ in open space. }
    \label{fig:order}
\end{figure}

\begin{figure}
    \centering
    \includegraphics[width = 8.5cm]{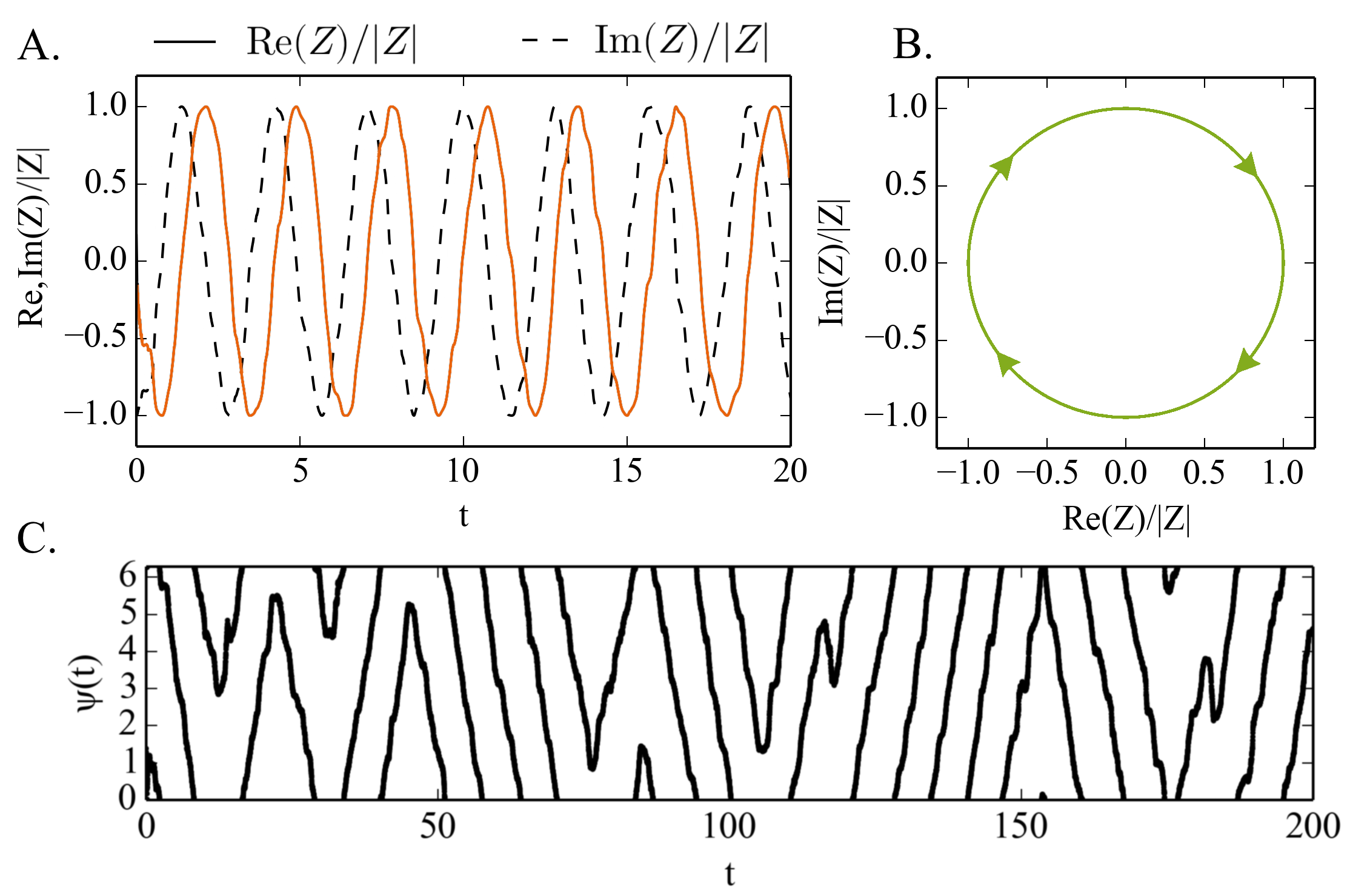}
    \caption{A,B) Behavior of the model for $\mathcal{D} = 30$ and $N_\text{int} = N/2$, $N = 200$. Oscillations in the two (normalized )order parameters $(\text{Re}(Z) /|Z|,\text{Im}(Z) /|Z|)$ signaling a non-zero collective angular velocity of $\psi(t)$, similar to the results of Bonilla \emph{et al} \cite{bonilla2019contrarian}. C) Time-series of the global direction of motion $\psi(t)$ for $\mathcal{D} = 7$, $N = 200$, $N_\text{int} = 0.3 N$. Stochastic sign changes in the global angular velocity $\dot \psi$ are observed. }
    \label{fig:Bonilla}
\end{figure}

Active particles have an intrinsic mobility, and move at a constant speed $v_0$ in the direction dictated by the orientation angle $\varphi_i(t)$ which changes in time as in Eq.~(\ref{eq:stochastic}). In dimensionless units set by a scale $\ell$, we have that $\dot{\mathbf{x}}_i = \text{Pe}_s \hat{e}_i(t)$ where $\text{Pe}_s = v_0 /(\gamma \ell)$ is the P\'{e}clet number of the microswimmer, and $\hat{e}_i(t) = [\cos \phi_i, \sin \phi_i] \equiv [e_x^i, e_y^i]$. Note that we have two model parameters $(\mathcal{D}$ and $\text{Pe}_s)$ that control internal and external properties of the system. The delay number says something about how quickly particles synchronize their orientation angle compared to how quickly they undergo SPRs. The P\'{e}clet number measures the persistence length in terms of some reference lengthscale. In open space, this scale can be taken to be the persistence length itself, and the P\'{e}clet number is just unity. Under confinement, a natural choice for $\ell$ is the system size, in which case the P\'{e}clet number becomes a tunable parameter measuring the strength of confinement relative to intrinsic dynamics. Our model is formally equivalent to a Kuramoto model of coupled oscillators on a random network driven by a Poissonian noise with a coupling to spatial dynamics. In the absence of any boundaries or obstacles that alter the particle's direction of motion, this is a one-way coupling; the angular dynamics feeds into the spatial dynamics, while the converse is not true. In the presence of confinement, the behavior is more complex, since collisions with boundaries may induce (approximately) instantaneous changes in the direction of motion. We investigate the role of SPRs on the collective dynamics both under confinement (fig. (\ref{fig:sketch} A)) and in open space (fig. (\ref{fig:sketch} B)).  

To quantify the ordered flocking states of the interacting active particles, we define two real order parameters $(r,\psi)$ or equivalently a complex order parameter $Z$ analogous to that of the phase synchronization in the Kuramoto model, 
\begin{equation} \label{eq:order}
Z(t) = r(t) e^{-i \psi(t)} = \frac{1}{n}\sum_{i=1}^n e^{-i \varphi_i(t)},    
\end{equation}
which may alternatively be expressed as $r^2 = \overline{e_x}^2 + \overline{e_y}^2$ and $\psi =  \tan^{-1}(\overline{e_y}/\overline{e_x})$, with the bar notation denoting averages over all particles. We see that $r$ is the mean-squared normalized velocity of the particles, while $\psi$ is their average direction of motion. In the disordered state where $r =0$, Eq. (\ref{eq:order}) implies that the orientation angles are uniformly distributed, while highly ordered states $r \approx 1$ happens when $\varphi_i \approx \psi$.

Valuable insights can be gained in the mean-field approximation of our model, corresponding to $N_\text{int} = n$. In the absence of confinement, the orientation $\phi$ has a probability density $\rho(\phi,t)$ that follows the Fokker-Plank equation
\begin{equation}\label{eq:MFT}
    \partial_t \rho(\phi) = \partial_\phi \left [ r \mathcal{D} \rho(\phi) \sin(\phi-\psi) \right] - \rho(\phi) + \rho(\phi+ \pi),
\end{equation}
where the first term is the same as in the mean-field Kuramoto model \cite{strogatz1991stability}, while the last two terms describe loss and gain terms correspond to Poissonian reversals. The order parameter from Eq.~(\ref{eq:order}) reduces to $Z = \langle \exp(- i \phi)\rangle$, with brackets indicating average over the probability distribution. Note that $Z$ is nothing but the first Fourier mode of the probability density. 

The transition from disordered to ordered states takes place at a critical value of the delay number $\mathcal{D}$. In the mean-field limit, this critical number is determined from the linear stability analysis of Eq.~(\ref{eq:MFT}) \cite{strogatz1991stability}. By linearizing Eq.~(\ref{eq:MFT}) around the disordered state, i.e. $r=0$ and $\rho(\phi) = 1/(2\pi)$, and using the fact that the complex order parameter is the first Fourier mode of the density, one can derive an evolution equation for the order parameter as $\dot r/r = \mathcal{D}/2 - 2$. Hence for the order parameter to grow, we need $\mathcal{D} > 4 \equiv \mathcal{D}_c$. In terms of dimension-full quantities the transition takes place when the mean waiting time between single-particle reversals is larger than four times the relaxation time of the interactions. Fig. (\ref{fig:order} A-B) shows the order parameter $Z(t)$ as function of time (panel A) and the dependence of its asymptotic value on the delay number both in the mean-field limit ($N_\text{int} = N$) and for a more sparse interaction network ($N_\text{int} \ll N$). We see a clear agreement with the predicted $\mathcal{D}_c = 4$ even beyond the mean field approximation for sufficiently large number of particles. As expected with decreasing the number of particles, the transition is less abrupt, but remains consistent with predictions.

Fig. (\ref{fig:order} C-E) shows representative center-of-mass trajectories for different $\mathcal{D}$, which below $\mathcal{D}_c$ results in random meanderings (panel C). Above this critical value (panels D and E) chiral collective motion develops. At intermediate values of $\mathcal{D}$, the chirality switches handedness spontaneously, as can also be seen from the time series of $\psi$ in Fig.~(\ref{fig:Bonilla} C). See movie 1 in the Supplemental Information for an example of this behavior. Fig.~(\ref{fig:Bonilla} A-B) shows the behavior for large $\mathcal{D}$, where the system locks into a state of fixed chirality, corresponding to a flocking state where the flocks direction of motion rotates with a fixed handedness. If we interpret the collective flock as a quasiparticle, the dynamics of this is reminiscent of that of a chiral active Brownian particle with stochastic chirality switching \cite{PhysRevE.103.052608}. It is important to note that the individual particles do not have an imposed chirality, and that the collective chiral behavior in the (partially) ordered state is an emergent property of the system originating in the Poissonian noise and finite-time interactions. Similar emergence of chiral macroscopic states have been observed in \textit{Myxococcus xanthus} bacterial colonies~\cite{balagam2021emergent}.

\begin{figure*}
    \centering
    \includegraphics[width = 17.2cm]{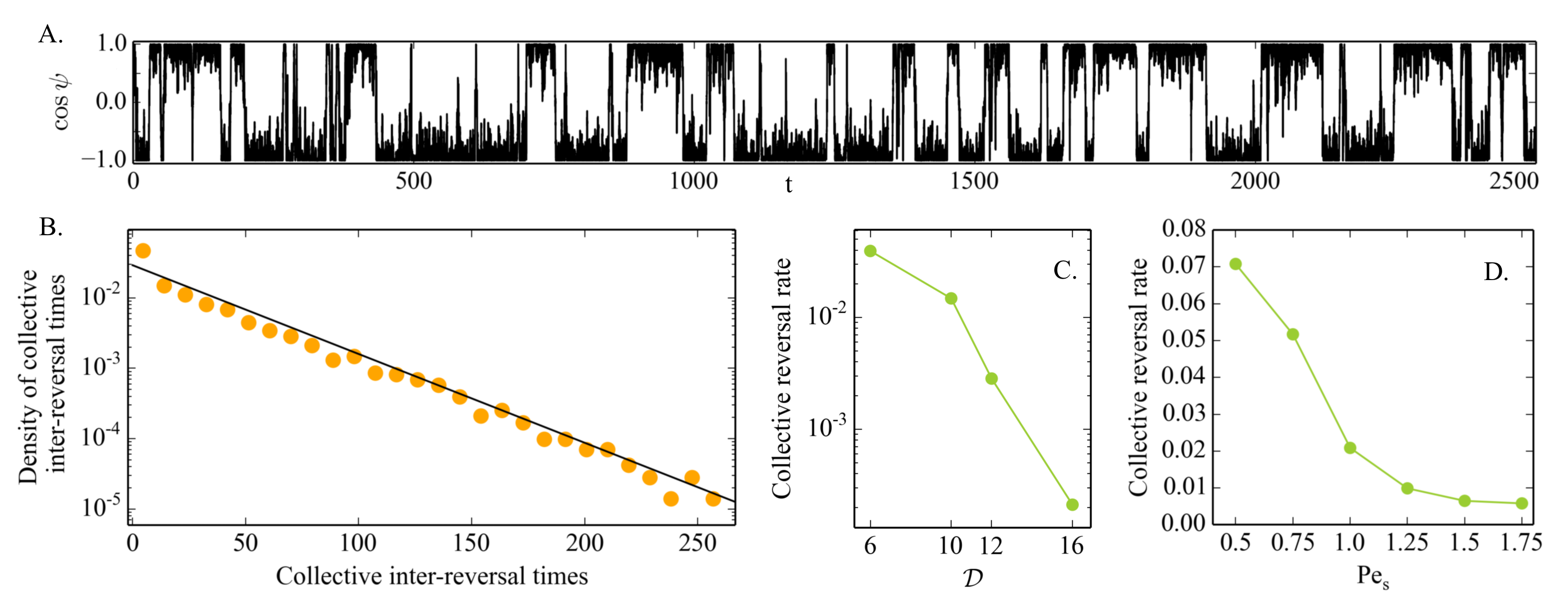}
    \caption{A) Channel order parameter $\cos\psi$ as a function of time, showing clear signs of reversals. Here $N = 500, N_\text{int} = 0.2 N, Pe = 1, \mathcal{D} = 7$, resulting in an order parameter $r \approx 0.32$. B) Collective waiting time density associated with a long timeseries as in A. C) Collective reversal rate as a function of the dimensionless number $\mathcal{D}$. As $\mathcal{D}$ increases, corresponding to increasing interaction strengths, the collective reversal rate decreases. For smaller values of $\mathcal{D}$ (and smaller values of the order parameter $r$) the reversals are indistinguishable from noise. D) In the limit of strong confinement ($\text{Pe}_s$ increasing) the collective reversal rate decreases as a consequence of the boundary interactions promoting ordered states.  }
    \label{fig:rev}
\end{figure*}

The mean-field approximation again provides us a qualitative understanding how the Poissonian noise and the alignment interaction influence the stability of the ordered state. We now linearize at the short-time dynamics of the mean-field orientation $\phi$ away from the perfectly ordered state, i.e. $\phi = \psi + \delta \phi$. The evolution of its average is $\partial_t \langle \phi \rangle = - r D \langle \sin(\phi-\psi) \rangle -  \pi$, where the last term comes from the Poissonian noise, such that the perturbation away from the order state evolves as 
\begin{equation}
    \partial_t \langle\delta \phi\rangle \approx - r \mathcal{D} \langle\delta \phi\rangle - \pi
\end{equation}
We see that the first term has a stabilizing contribution, while the second term corresponding to Poissonian reversals however is destabilizing the ordered state. If instead, we consider the case relevant when a single particle reverses its direction of motion and moves opposite to the flock $\phi = \psi + \pi +  \delta \phi$, the linearized dynamics reads 
\begin{equation}
    \partial_t \langle\delta \phi\rangle \approx  r \mathcal{D} \langle\delta \phi\rangle - \pi
\end{equation}
Small deviations are now destabilized both by alignment interactions and Poissonian reversals.  

This destabilizing effect is crucial to understanding the collective motion in the system. If a particle randomly reverses its direction of motion, any small perturbation causes either a clockwise or anti-clockwise drift towards the global direction of motion. This would also affect the evolution of the global orientation $\psi$. When several SPR can take place over the same time window and strong correlations are present, one particle's random angular drift in orientation will induce quick adjustments of the others, resulting in a biased macroscopic angular drift. In this regime, we expect $\psi$ to evolve similar to that of dichotomous diffusion, where the angular velocity $\dot \psi$ behaves like a telegraph noise. This is indeed what is observed in Fig. (\ref{fig:Bonilla} C). In the limit of high $\mathcal{D}$, the system gets locked into a chiral state with a fixed handedness (see Fig. (\ref{fig:order} E) or Fig. (\ref{fig:Bonilla} A)), and displays little spatial dynamics. Interestingly, the reversals on the microscopic scale do not lead to sudden collective reversals of the global direction of motion as one might expect. We attribute this to the lack of a symmetry breaking in the system; the system preserves its a rotational invariance. Rather than undergoing spontaneous macroscopic reversals $\psi \to \psi + \pi$ the system explores all the equivalent macrostates which can be related through the rotational symmetry. 

One way to break the rotational symmetry is to include a non-rotationally invariant confinement. When such confinement is introduced, the story becomes more subtle. We consider an infinite channel geometry that is periodic in $x$ and has a finite extent in the $y$-direction. The channel's width is used to set the relevant length scale $\ell$ which allows us to tune the Peclet number away from unity. First, let us consider the expected ordered states. Because of the confinement's geometry, there can be no average motion in the vertical direction, and hence $\overline{e_y} \approx 0$. This is of course not true exactly at every time $t$, but, on a slightly coarser time scale, the fluctuations average to zero. This immediately implies from Eq.~(\ref{eq:order}) that $r \sin\psi = 0$. For disordered states ($r=0$), this is trivially satisfied, while for states with some degree of order ($r > 0$), this can only hold if the global direction of motion satisfies $\psi = n \pi$, implying that $\cos \psi = \overline{e_x}/r = \pm 1$. Hence in a channel confinement, it will be useful to quantify the order in the system by the order parameters $(r,\cos \psi)$, where the latter dictates left $(-1)$ or right-moving $(+1)$ collective motion. The rotational symmetry in open space where $\psi$ takes any values $\psi \in (0,2\pi)$ is now broken, and the remaining ordered states are now connected instead by a discrete symmetry. The system spontaneously transitions between the left- and right-moving states, similar to what has been observed in some systems of collective animal behavior confined to an annulus  \cite{chen2021coordinating,yates2009inherent,buhl2006disorder}. Fig.~(\ref{fig:rev} A) shows the order parameter $\cos\psi$ undergoing reversals like a telegraphic noise. To gain further insights into the statistical nature of these reversals, we study the statistics of the inter-reversal time, or the waiting time between collective reversals. Fig. ~(\ref{fig:rev} B) shows a clear exponential distribution of the inter-reversal times, which points to the fact there is the collective reversals are not correlated with one another. Furthermore, Fig.~(\ref{fig:rev} C) shows that the characteristic reversal rate decreases with increasing the delay number $\mathcal D$, hence this macroscopic reversal behavior tends to fade away into noise for sufficiently frequent SPRs. At the opposite end of the spectrum, when sufficiently strong interactions make it difficult to destabilize the ordered state, such that macroscopic reversals become extremely rare events. A nontrivial dependence is also observed with $\text{Pe}_s $. For small values of the Peclet number the system becomes effectively unconfined and it is not possible to identify collective reversals between the two macroscopic states (left- and right-moving flocks). As the Peclet number increases the particles more strongly interact with the solid boundaries. Such interactions promote ordered states with $\cos\psi = \pm 1$ and hence the flocks will be locked into their collective direction of motion. As with the delay number, the strongest reversal phenomena are observed at intermediate values (See Movie 2 in the Supplemental Information for an example of collective directional reversal in confinement).

In summary, we studied a minimal model of active particles with topological alignment interactions and random single particle reversals and found surprising emergent behavior both in open spaces and under symmetry-breaking confinement. In open space, where the rotational symmetry is preserved, we predict that there is a critical delay number $\mathcal{D}_c=4$ above which the system transitions spontaneously from a disordered state to emergent collective chiral states, resulting from the combined effect of stochastic SPR's and a finite timescale of interactions. When the active particles are confined to a channel geometry, collective spontaneous switching between left- and right-going currents are observed. The collective directional switching resemble a telegraphic noise and has a typical rate that decays quickly with both $\mathcal{D}$ and $\text{Pe}_s$. Hence we expect that such reversal phenomena are not as easily observed in the traditional Vicsek model, since in that case the alignment timescale is zero and hence $\mathcal{D}\to \infty$. Similar phenomena are however expected to be present in biological systems where interactions have a non-zero timescale, for example in systems with sensorial delay or rotational inertia.

\begin{acknowledgements}
 K.S.O acknowledges support from the Nordita Fellowship program. L.A and E.G.F acknowledge support from the Research Council of Norway through the Center of Excellence funding scheme, Project No. 262644 (PoreLab). 
\end{acknowledgements}

 %\bibliographystyle{apsrev4-2}
 %\bibliography{OlsenEtAl.bib}


\end{document}